\documentclass[aps,pra,preprint,groupedaddress,showpacs]{revtex4}

\usepackage{graphicx}% Include figure files
\usepackage{dcolumn}% Align table columns on decimal point
\usepackage{bm}% bold math

\newcommand{\bra}[1]{\langle #1 |}
\newcommand{\ket}[1]{| #1 \rangle}

\newcommand{\meanI}[1]{\langle #1 \rangle}
\newcommand{\var}[2]{\meanI{(\Delta \hat{#1}_{#2})^{2}}}

\begin{document}

%******************[ Title page includes the authors and abstract ]*******************%
\title{Deterministic Secure Communications using Two-Mode Squeezed States}
\author{Alberto M. Marino}
\email{marino@optics.rochester.edu}
\author{C. R. \surname{Stroud},~Jr.}
\affiliation{The Institute of Optics, University of Rochester,
Rochester, New York 14627, USA}
%\date{\today}

\begin{abstract}
We propose a scheme for quantum cryptography that uses the squeezing
phase of a two-mode squeezed state to transmit information securely
between two parties. The basic principle behind this scheme is the
fact that each mode of the squeezed field by itself does not contain
any information regarding the squeezing phase.  The squeezing phase
can only be obtained through a joint measurement of the two modes.
This, combined with the fact that it is possible to perform remote
squeezing measurements, makes it possible to implement a secure
quantum communication scheme in which a deterministic signal can be
transmitted directly between two parties while the encryption is
done automatically by the quantum correlations present in the
two-mode squeezed state.
\end{abstract}


\maketitle
%******************************[ Intro to the topic ]*************************************%

One of the most interesting applications to have emerged from the
recent growth in quantum optics is quantum cryptography. This field
started with the pioneering work of Bennet and Brassard
\cite{Bennett84}. The basic idea is to use unique quantum mechanical
properties, such as entanglement and superposition of states, in
order to transmit a signal securely between two parties. These
fundamental quantum mechanical properties allow, in principle, to
detect the presence of an eavesdropper with certainty, thus making
it possible to determine if a channel is secure. The basic mechanism
used in quantum cryptography for the detection of an eavesdropper is
that any measurement inevitably disturbs the state of the system.

All of the initial quantum cryptography schemes were based either on
single photons \cite{Bennett84} or entangled photon pairs
\cite{Ekert91}. As a result, their experimental implementation is
limited due to the lack of good single photon sources as well as
efficient single photon detectors. In order to overcome these
limitations there have been some recent proposals that extend the
ideas of  discrete quantum cryptography into the domain of
continuous variables (CV)
\cite{Ralph00a,Reid00,Pereira00,Hillery00,Gottesman01,Silberhorn02a,Funk02}.
For these new schemes, the two-mode squeezed state (TMSS) has proven
to be the basic source of continuous variable entanglement. These
quantum states of the field are routinely generated, the most common
method being the optical parametric oscillator (OPO)
\cite{Wu87,Lam99}, and can be detected with high efficiency through
the use of homodyne detection.

In this paper we propose a new CV quantum cryptography scheme that
uses the squeezing phase of a TMSS in order to transmit information
between two parties. The scheme is based on the facts that each mode
of a TMSS by itself does not contain any information regarding the
squeezing phase, and that it is possible to perform remote squeezing
measurements. As opposed to most previous schemes, the scheme
presented here allows for the secure transmission of a deterministic
signal. It thus allows for the secure transmission of an encryption
key as well as the possibility of sending a message directly. In the
case in which a message is transmitted directly, the encryption is
done automatically by the quantum correlations present between the
modes of a TMSS. As with any other quantum communication protocol,
it is necessary to verify the security of the quantum channel used.
This can be done, as well, with the help of the quantum correlations
present in a TMSS \cite{Reid89,Barnett91}.

We start by looking at the relevant properties of a TMSS.  The
two-mode squeezed state is defined according to \cite{Loudon87}
\begin{equation}
    \ket{\alpha,\beta;\zeta}=\hat{D}_{1}(\alpha)\hat{D}_{2}(\beta)\hat{S}_{12}(\zeta)\ket{0,0},
\end{equation}
where $\hat{D}(\alpha)=\exp(\alpha\hat{a}^{\dag}-\alpha^{*}\hat{a})$
is the displacement operator,
$\hat{S}_{12}(\zeta)=\exp(\zeta^{*}\hat{a}_{1}\hat{a}_{2}-\zeta\hat{a}_{1}^{\dag}\hat{a}_{2}^{\dag})$
is the two-mode squeezing operator, and $\zeta=se^{i\theta}$ is the
squeezing parameter. The quantum correlations present in a TMSS
become more evident when it is written in terms of number states.
Such an expansion can be shown to be given by \cite{Barnett87}
\begin{equation}
\label{TMSS_number}
    \ket{\alpha,\beta;\zeta}=\frac{1}{\cosh
    s}\sum_{n=0}^{\infty}(-e^{i\theta}\tanh
    s)^{n}\ket{\alpha;n}\ket{\beta;n},
\end{equation}
where $s$ is the degree of squeezing, $\theta$ is the squeezing
phase, and $\ket{\alpha;n}=\hat{D}(\alpha)\ket{n}$ is the displaced
number state. As we can see from Eq.~(\ref{TMSS_number}), there are
perfect quantum correlations between the displaced number states of
the two modes.

We now investigate the properties of the individual modes through
the reduced density matrices, which from Eq.~(\ref{TMSS_number}) can
be shown to be given by
\begin{eqnarray}
    \hat{\rho}_{1} & = & \frac{1}{\cosh^{2} s}\sum_{n=0}^{\infty}(\tanh
    s)^{2n} \ket{\alpha;n}\bra{\alpha;n} \nonumber\\
    \hat{\rho}_{2} & = & \frac{1}{\cosh^{2} s}\sum_{n=0}^{\infty}(\tanh
    s)^{2n} \ket{\beta;n}\bra{\beta;n}. \label{TMSS_reduced_b}
\end{eqnarray}
The main thing to note from Eq.~(\ref{TMSS_reduced_b}) is that each
mode by itself carries no information about the squeezing phase,
only about the degree of squeezing.  As a result, if one looks at
the noise of an individual mode using homodyne detection, it will be
phase independent. It is necessary to make a combined measurement of
the two modes in order to extract the squeezing phase. It is then
possible to take advantage of this property of a TMSS and use the
squeezing phase to encode information. In fact, each mode by itself
has more noise than a coherent state; however, due to the fact that
the noise in the two modes is correlated it is possible to get the
noise below that of a coherent state when making a joint
measurement. This fact was used by Pereira \textit{et
al.}~\cite{Pereira00} in order to hide a classical message in the
excess noise of each mode and then use the quantum correlations
between the modes to extract the message.

\begin{figure}[ht]
    \includegraphics{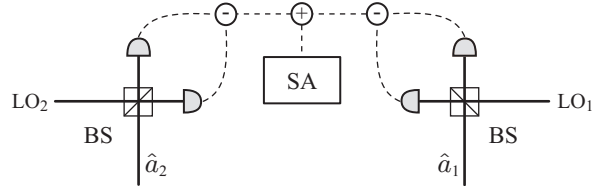}
    \caption{\label{remoteBLO} Detection scheme for remote squeezing measurements.
    A local heterodyne measurement of each mode is performed, after
    which the resulting photocurrents are combined to obtain a
    complete characterization of the two-mode squeezed state. Operators $\hat{a}_{1}$ and
    $\hat{a}_{2}$ represent the two modes of the squeezed state. \textit{Notation}: LO = local oscillator; BS = beam
    splitter; SA = spectrum analyzer.}
\end{figure}

In order to take advantage of the fact that the squeezing phase can
only be obtained from a combined measurement of both modes of the
squeezed field, it is necessary to use a measurement technique that
allows us to perform a remote squeezing measurement. This can be
done using the scheme shown in Fig.~\ref{remoteBLO} \cite{Ou92}.
With this detection scheme it is possible to perform local balanced
homodyne measurements on each of the modes of the TMSS and then
combine the resulting measurements in order to extract the squeezing
phase information. Once the two measurements are combined, the
variance of the resulting signal can be shown to be of the form
\begin{equation}
\label{remoteBLOnoise}
    \var{n}{12}=2|E_{LO}|^2\left[e^{-2s}\cos^{2}\left(\frac{\chi_1+\chi_2-\theta}{2}\right)
    +e^{2s}\sin^{2}\left(\frac{\chi_1+\chi_2-\theta}{2}\right)\right],
\end{equation}
where $E_{LO}$ is the amplitude of the local oscillator (LO),
$\chi_1$ is the phase of LO1, and $\chi_2$ is the phase of LO2. In
deriving this expression it has been assumed that the amplitude of
the two local oscillators is the same. As can be seen from
Eq.~(\ref{remoteBLOnoise}), it is possible to change the quadrature
that is measured by changing the phase of either LO.

\begin{figure}[ht]
    \includegraphics{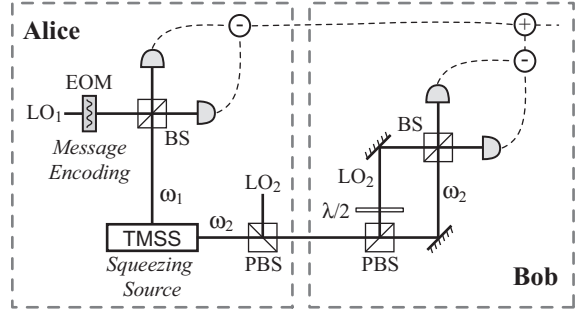}
    \caption{\label{securecomm}Proposed quantum cryptography scheme.
    The squeezing phase of the TMSS is used to transmit information between Alice and Bob.
    The information is encoded through the modulation of the local oscillator used by Alice to perform
    her homodyne measurement. The information is decoded by Bob by
    combining his and Alice's measurement results. \textit{Notation}:
    TMSS = two-mode squeezed state; LO = local oscillator; EOM = electro-optical modulator; BS = beam
    splitter; PBS = polarizing beam splitter; $\lambda$/2 = half-wave plate.}
\end{figure}

Using these two elements it is possible to implement a new quantum
cryptography scheme that uses the squeezing phase of a TMSS to
transmit information between two parties, Alice and Bob. The basic
communication scheme is shown in Fig.~\ref{securecomm}.
Alice uses a squeezed light source, such as an OPO, to generate a
TMSS and sends one of the modes to Bob over a quantum channel while
keeping the other mode. In order for Bob to perform his homodyne
measurements, he needs to have a LO which is phase coherent with the
mode he receives. Alice can also send a coherent state LO to Bob by
combining it with the squeezed beam with a polarizing beam splitter
(PBS). Bob then splits the LO and the squeezed mode using another
PBS and performs a homodyne measurement. When performing his
measurements, Bob keeps the phase of the LO constant. As can be seen
from the reduced density matrix of the TMSS,
Eq.~(\ref{TMSS_reduced_b}), Bob will have gained no information,
except for the degree of squeezing, by performing this measurement.
Which means that if an eavesdropper, Eve, were to intercept the mode
sent to Bob, she would not be able to extract any information
either.

As can be seen from Eq.~(\ref{remoteBLOnoise}), the noise of the
combined homodyne measurements of the modes of the TMSS can be
switched between the minimum and maximum noise levels by performing
a $\pi$ phase shift on the LO of either Alice or Bob. Thus, in order
to transmit the signal, Alice encodes the information on her part of
the TMSS by changing the phase of the LO she uses to perform her
homodyne measurements. This can easily be done with the help of an
electro-optic modulator, thus allowing for a high speed encoding of
the signal. Once Alice has encoded the information on her mode by
making the homodyne measurements and Bob has finished making his set
of homodyne measurements, she can send her measurement results to
Bob over a public channel. As is the case with the mode sent to Bob,
the measurement results obtained by Alice will contain no
information on the encoded signal. It is necessary to combine the
two measurements in order to retrieve the signal. Once Bob receives
the measurements from Alice he can combine both his and Alice's
measurements to decode the signal. Since Bob keeps the phase of his
LO constant, once he combines both measurements, he will see a
signal whose variance will be changing between two different levels
which represent the information encoded by Alice.

As with any other quantum key distribution (QKD) scheme, it is
necessary to verify the security of the quantum channel used to send
Bob his part of the TMSS. In the proposed scheme described above,
this can be done through the quantum correlations present in the
TMSS. Since Bob does not need to change the phase of his LO when
performing his measurements, all the information he decodes after
combining both his and Alice's measurements can be used both for
generating a cryptographic key and verifying the security of the
channel. A complete security analysis of the proposed scheme is
beyond the scope of this paper. Instead, we focus on the usual
attack strategies for continuous variables \cite{Ralph00a,Ralph00b},
such as intercept and resend and partial interception of the mode
sent to Bob.

Once the measurement is performed by Alice, the resulting
photocurrent can be treated as a classical signal \cite{Bachor}.
Thus, as long as the electronic noise of the components used is low
enough, any division or amplification of this classical signal will
not affect the final result. In the following analysis we thus
consider the worst case scenario, which corresponds to one in which
Eve can make a perfect copy of the information sent through the
public channel without modifying it.

The easiest form of attack is the intercept and resend strategy.  In
this case Eve intercepts the mode sent to Bob.  As can be seen from
Eq.~(\ref{TMSS_reduced_b}), Eve will have gained no information on
the squeezing phase from this mode and thus no information on the
transmitted signal.  Since in principle it is possible to extract
the degree of squeezing present in the TMSS from the measurement of
a single mode, Eve could resend a mode of a TMSS with the same
degree of squeezing to Bob. However, the new mode sent out by Eve
will not be entangled with the mode retained by Alice.  Thus, when
Bob combines his and Alice's measurement results the combined signal
will contain more noise than a coherent state and will be phase
independent. That is, there will be no squeezing and the modulation
of the variance will not be present.  Thus, this type of attack can
easily be detected.

\begin{figure}[ht]
    \includegraphics{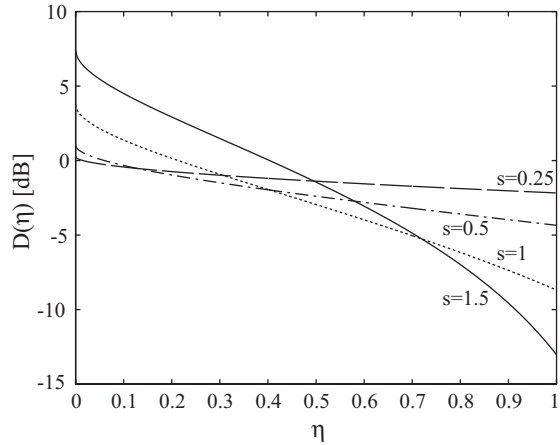}
    \caption{\label{degreesqueeze} Degree of squeezing measured by Bob after combining his and Alice's measurements
    as a function of the losses, 1-$\eta$, in the mode sent to Bob.}
\end{figure}

The other usual strategy of attack is partial interception.  In this
case Eve splits part of the signal and use this together with the
classical signal to gain information on the transmitted signal. To
see the effect this type of attack would have on the degree of
squeezing we need to look at the case in which only the mode sent to
Bob experiences losses.  If we assume that Eve intercepts a portion
$1-\eta$ of the mode sent to Bob, then the variance of the combined
signal takes the form
\begin{eqnarray}
    \label{TMSSlosssinglemode}
    \var{n}{12}&=&|E_{LO}|^2\left\{(1-\eta)+e^{-2s}\left[\frac{1+\eta}{2}+\sqrt{\eta}\cos(\chi_{1}+\chi_{2}-\theta)\right]\right.\nonumber\\
    &&\left.+e^{2s}\left[\frac{1+\eta}{2}-\sqrt{\eta}\cos(\chi_{1}+\chi_{2}-\theta)\right]\right\}.
\end{eqnarray}
In the case of a coherent state Eq.~(\ref{TMSSlosssinglemode}) can
be shown to be phase independent and to reduce to the expected level
of $2|E_{LO}|^2$.  In order to see the effect of partial
interception on the amount of squeezing that is measured, we look at
the measured degree of squeezing, which is defined according to
\begin{equation}
    D(\eta)=10\log_{10}\left[\frac{\var{n}{12}_{min}}{\var{n}{12}_{cs}}\right],
\end{equation}
where $\var{n}{12}_{min}$ is the minimum variance of the squeezed
state and $\var{n}{12}_{cs}$ represents the variance of a coherent
state. This quantity is plotted for different initial degrees of
squeezing in Fig.~\ref{degreesqueeze}.
As can be seen from this figure any loss will result in a reduction
of the degree of squeezing, as expected.  The rate at which this
happens becomes larger for a higher initial degree of squeezing,
which means that the losses become more evident. Thus the ability to
detect any eavesdropping increases as the inital squeezing parameter
$s$ increases. An interesting thing to notice from
Fig.~\ref{degreesqueeze} is that as $\eta\rightarrow 0$ the noise
goes above that of the coherent state. The reason for this is that
each mode by itself has excess noise, so that if there is loss in
only one of the modes then the cancelation of the excess noise is
not perfect, leading to more noise than a coherent state for large
losses.

\begin{figure}[ht]
    \includegraphics{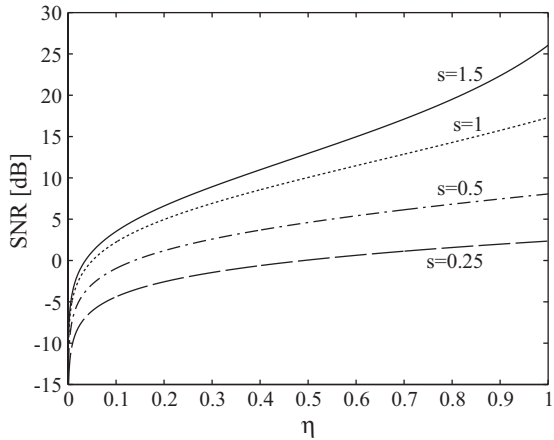}
    \caption{\label{SNRfig} Signal-to-noise ratio of quantum
    cryptography scheme
    as a function of the losses, 1-$\eta$, present in the mode of the TMSS sent to Bob. }
\end{figure}

If the minimum and maximum values of the variance,
Eq.~(\ref{TMSSlosssinglemode}), are used for the transmission of the
signal, then the signal-to-noise ratio (SNR) is given by
\begin{eqnarray}
    \label{SNR}
    SNR&=&\frac{\var{n}{12}_{max}-\var{n}{12}_{min}}{\var{n}{12}_{min}}\nonumber\\
    &=&
    \frac{4\sqrt{\eta}\sinh
    2s}{(1-\eta)+e^{-2s}[(1+\eta)/2+\sqrt{\eta}]+e^{2s}[(1+\eta)/2-\sqrt{\eta}]}.
\end{eqnarray}
The effect of losses on the SNR can be seen in Fig.~\ref{SNRfig}. We
again see that a higher degree of squeezing will lead to a larger
change in SNR for a given amount of loss.
From these results we have, for example, that for $s=1$, which
corresponds to squeezing of around 8.5~dB, a reduction of the amount
of squeezing measured and of the SNR of about 1~dB requires the
interception of only 7\% of the mode.  Thus making it a very
sensitive way of detecting any interception of part of the mode sent
to Bob.

Eve could instead use a quantum attack, such as an optical tap
\cite{Poizat93} or a quantum non-demolition measurement
\cite{Grangier98} in order to gain some extra information without
being detected. However, this type of attack would lead to increased
noise in the quadrature conjugate to the quadrature that is
measured. Since the proposed scheme effectively measures both
quadratures, this type of attack can also be identified by
monitoring the quantum correlations of the TMSS. By also modulating
the squeezing phase $\theta$, in addition to Alice's LO, an extra
degree of security can be added for the quantum attacks described
above. As can be seen from Eq.~(\ref{remoteBLOnoise}), as long as
the quantity $\chi_{1}-\theta$ is changed by a factor of $\pi$ the
scheme presented above will not be modified.

In practice, the degree of squeezing or correlations for a given
quantum channel can be established. Thus, once the losses in the
quantum channel have been accounted for, both the degree of
squeezing and the SNR can be used to verify if unauthorized access
to the quantum channel has occurred.

As opposed to most previous schemes, which rely on both Alice and
Bob making random measurements of the quadratures of the field, the
scheme presented here does not require Bob to actively select
between different measurements.  Only a single beam, Alice's LO,
needs to be phase modulated in order to encode the signal. As a
result, a higher transmission rate can be achieved due to the fact
that all the information received by Bob can be used both to
generate the encryption key and to verify the security of the
quantum channel through the quantum correlations present in the
TMSS.

The fact that a deterministic signal can be sent with our scheme
opens up the possibility of using it for the secure transmission of
a message. This allows the bypassing of the encryption-decryption
process, thus making the communication process more efficient while
maintaining the security of the proposed continuous variable QKD
scheme. When the scheme is used for the transmission of a
cryptographic key, if the presence of an eavesdropper is detected
then the whole key can be thrown away and a new one generated.
However, this is not the case when a message is sent. In this case,
in order to verify the security of the channel, Alice needs to
insert check bits at random times throughout the message. Since the
check bits are inserted at random times, Eve will not know when to
perform a measurement in order to intercept the message but not the
check bits.  As a result, the eavesdropping  will affect both the
message and the check bits. Once Bob measures the mode he received,
Alice sends her measurements results on the check bits and tells Bob
which time slots correspond to these bits.  Bob can then use the
information obtained from combining his and Alice's results for the
check bits in order to verify the security of the quantum channel as
described above.  Once the security of the channel has been
verified, Alice can send her measurement results for the actual
message to Bob, who just needs to combine them with his results in
order to decode the message.  If the presence of an eavesdropper is
detected, a different quantum channel needs to be used and the
processes repeated.

In conclusion, we have presented a new quantum cryptography scheme
that relies on the squeezing phase of a TMSS for the secure
transmission of information. The security of the quantum channel
used can be verified with the help of the quantum correlations
present in the squeezed field, since the presence of an eavesdropper
will result in a decrease in the degree of squeezing measured as
well as an increase in the errors in the transmitted signal. Apart
form the usual application of quantum cryptographic systems for the
transmission of encryption keys, this new scheme allows for the
transmission of a deterministic signal between two parties, thus
opening the possibility for direct transmission of messages.

\end{document}